\documentclass[11pt]{article}

\usepackage[utf8]{inputenc}
\usepackage{graphicx,amsmath,amsfonts,amssymb,fullpage,natbib}

\usepackage[T1]{fontenc}

\title{\textbf{The molecular memory code and synaptic plasticity: a synthesis}}
\author{Samuel J. Gershman \\
Department of Psychology and Center for Brain Science, Harvard University\\
Center for Brains, Minds and Machines, MIT \\
Correspondence: gershman@fas.harvard.edu}

\begin{document}

\maketitle

\begin{abstract}
    The most widely accepted view of memory in the brain holds that synapses are the storage sites of memory, and that memories are formed through associative modification of synapses. This view has been challenged on conceptual and empirical grounds. As an alternative, it has been proposed that molecules within the cell body are the storage sites of memory, and that memories are formed through biochemical operations on these molecules. This paper proposes a synthesis of these two views, grounded in a computational theory of memory. Synapses are conceived as storage sites for the parameters of an approximate posterior probability distribution over latent causes. Intracellular molecules are conceived as storage sites for the parameters of a generative model. The theory stipulates how these two components work together as part of an integrated algorithm for learning and inference.
\end{abstract}

\newpage



\begin{quote}
    ``The lack of interest of neurophysiologists in the macromolecular theory of memory can be accounted for by recognizing that the theory, whether true or false, is clearly premature. There is no chain of reasonable inferences by means of which our present, albeit highly imperfect, view of the functional organization of the brain can be reconciled with the possibility of its acquiring, storing and retrieving nervous information by encoding such information in molecules of nucleic acid or protein.''
    \\ \\
    \citet{stent72}, \emph{Prematurity and uniqueness in scientific discovery}
\end{quote}

\section{Introduction}

Understanding the biological basis of memory is one of the most profound puzzles in neuroscience. Solving this puzzle requires answers to two questions: (i) What is the \emph{content} of memory, and (ii) what is the \emph{structure} of memory? Content refers to the information encoded in memory. Structure refers to the code that maps information into a physical trace (or ``engram''). There are standard textbook answers to these questions: The content of memory consists of associations between events; the code is a mapping from associations to synaptic strengths \citep{martin00,takeuchi14}.

There are major problems with these textbook answers, which I will summarize below. An alternative hypothesis, originally proposed in the mid 20th century \citep[see][for a review]{gaito76} and recently revived \citep{gallistel17,abraham19}, postulates that memories are encoded in an intracellular molecular substrate. The precise nature of this substrate is speculative and a matter of debate. Gallistel has argued that the content of memory consists of ``facts'' (explicit representations of variables in the environment) rather than associations \citep{gallistel11,gallistel17,gallistel21}. These facts are read out by the spiking activity of neurons (presumably mediated by some biochemical process within the cell) and thereby made computationally accessible to downstream neurons. Synaptic plasticity, on this view, plays no role in memory storage.

As we will see, the molecular hypothesis has many merits. But if it turns out to be true, we are left with a residual puzzle: What is the function of synaptic plasticity? Note that Gallistel never denies the \emph{existence} of synaptic plasticity, only its functional implications. However, it would be odd if the brain carried out functionally inert computations that may demand on the order of $10\%$ of the total cortical metabolic rate \citep{karbowski19}. Moreover, synaptic plasticity is \emph{not} functionally inert, since it appears to be causally related to learned behavior \citep[reviewed by][and discussed further below]{takeuchi14}.

I will develop new answers to these questions, starting with a distinction between two kinds of memory, one synaptic and one intracellular, that serve different computational functions. Following Gallistel, I propose that an intracellular molecular mechanism stores ``facts''---in particular, the parameters of a \emph{generative model} (a concept I elaborate below). To do some useful computation with these facts (e.g., perception, prediction, action), the brain needs to infer the latent causes of sensory observations by inverting the generative model. This is generally intractable for complex generative models, but can be approximated using a parametrized mapping from observations to a distribution over latent causes---the \emph{inference model}.\footnote{The inference model is sometimes referred to as a \emph{recognition model} \citep{dayan95,hinton94,kingma13} or \emph{inference network} \citep{mnih14,rezende15}. In the machine learning literature, the inference model is typically implemented as a deep neural network.} I hypothesize that: (i) this inference model is implemented by spiking activity in a network of neurons; (ii) its parameters are stored at the synapse; and (iii) synaptic plasticity updates these parameters to optimize the inference model.

Importantly, the generative model and the inference model fit together as part of a single optimization problem, namely minimizing free energy. This is already a familiar idea in theoretical neuroscience \citep[e.g.,.][]{friston06,friston10,gershman19}. The most biologically detailed implementations of free energy minimization rest upon a predictive coding scheme in which feedforward pathways (neurons in superficial cortical layers) convey prediction errors and feedback pathways (neurons in deep cortical layers) convey predictions \citep{bastos12,bogacz17,friston05}. Both generative and inference parameters are learned via Hebbian synaptic plasticity rules. Thus, the main departure explored here is the idea that the generative model is implemented by intracellular computations, and learning the parameters of the generative model is implemented by molecular plasticity within cells rather than at the synapse.\footnote{It is also worth noting that predictive coding can only be applied to a restricted class of models, and also entails a Gaussian approximation of the posterior \citep[see][for further discussion]{gershman19}.}

To set the stage for my theory, I begin by surveying the empirical and conceptual issues with synaptic plasticity as a memory mechanism. This motivates a consideration of alternative (or complementary) mechanisms, in particular those based on intracellular molecular substrates. To understand how these synaptic and intracellular mechanisms fit together, I embark on an abstract theoretical treatment of the optimization problem facing the brain. I show how the optimization problem can be tractably (albeit approximately) solved, and how this solution can be implemented in a biologically plausible system. Finally, I discuss how this proposal reconciles the divergent views of memory in the brain.

\section{Empirical and conceptual issues}

How did we get to the current consensus that synaptic plasticity is the basis of memory formation and storage? While there exists a massive literature on synaptic plasticity from physiological, cellular, and molecular perspectives, here I will focus on the most compelling evidence that synaptic plasticity plays a role in the behavioral expression of memory. I will then critically evaluate this body of evidence.

\subsection{Synaptic plasticity and memory}

The classical way to study synaptic plasticity is to stimulate a presynaptic axon with a high-frequency tetanus and measure the size of evoked excitatory postsynaptic potentials (EPSPs). \citet{bliss73} first reported that this procedure resulted in a long-lasting increase in EPSP amplitude---a phenomenon known as \emph{long-term potentiation} (LTP). Later it was discovered that low-frequency stimulation reduces EPSP amplitude \citep{dudek92}---a phenomenon known as \emph{long term depression} (LTD). These findings were remarkable because they fit conveniently into an associative conceptualization of memory that dates back to Plato and Aristotle, and which was later championed by the British empiricist philosophers. In the 20th century, associative memory was propelled to the forefront of experimental psychology. Donald Hebb is usually credited with the hypothesis that the strengthening of connections between neurons might be the basis of memory formation \citep[``neurons that fire together, wire together'';][]{hebb49}, though William James proposed essentially the same idea (in terms of ``organic materials'' rather than neurons) more than half a century earlier \citep{james90}.

By now, these ideas have become so entrenched that it's difficult for neuroscientists to talk about memory without association. It has become a matter of \emph{definition} rather than \emph{theory}. For the moment, I will defer a critique of this definitional assumption. Taking it at face value, what is the evidence that LTP and LTD are causally responsible for memory as measured behaviorally? Few studies have addressed this question directly, despite our wealth of knowledge about LTP and LTD.

Some early attempts to address this question interfered with various cellular components that were considered to be crucial for synaptic plasticity. For example, NMDA receptors are necessary for LTP, and antagonizing these receptors in the hippocampus impairs learning in a hippocampus-dependent place learning task \citep[the Morris water maze;][]{morris86}. Unfortunately, this study, and many similar ones, suffers from a logical flaw: NMDA receptors (like many other components involved in synaptic plasticity) play multiple roles in the cell, so we do not know if the impairment is due specifically to blocked plasticity or to some other dysfunction in the cell \citep{shors97,nicoll17}. Moreover, subsequent work showed that animals can produce intact performance in the Morris water maze despite NMDA antagonism, as long as they are given some pretraining \citep{bannerman95,saucier95}. Thus, NMDA receptors do not appear to be a necessary component of learning, at least under certain conditions. I will discuss other examples of dissociations between LTP and learning in a later section.

Other studies have asked whether changes in synaptic strength are accompanied by changes in behavior. \citet{whitlock06} showed that inhibitory avoidance learning produced LTP at hippocampal synapses, accompanied by trafficking of AMPA receptors to the postsynaptic membrane. In the same spirit, \citet{rogan97} showed that fear conditioning produces LTP in the lateral amygdala, paralleling changes in cue-evoked fear responses \citep[see also][]{mckernan97}. Using an auditory frequency discrimination task, \citet{xiong15} showed that corticostriatal synapses were potentiated specifically for cortical neurons tuned to frequencies associated with reward, paralleling changes in discrimination performance.

Later studies provided causal evidence that trafficking of AMPA receptors to the postsynaptic membrane of lateral amygdala neurons is necessary for acquiring a fear memory \citep{rumpel05}, and endocytosis of AMPA receptors is necessary for the extinction of fear memory \citep{kim07,clem10}. Using optogenetic induction of LTP and LTD at amygdala synapses, \citet{nabavi14} were able to create, inactivate, and reactivate a fear memory. In motor cortex, optical erasure of dendritic spines formed during learning caused a selective disruption of the corresponding memory \citep{hayashi15}.

While these studies are suggestive, they suffer from another logical flaw: they do not tell us whether synaptic plasticity is necessary for the \emph{storage} of memory or for its \emph{expression}. If memories were stored in an intracellular format but required synapses to be expressed behaviorally (e.g., through conditioned responding), then changes in synapses should track changes in behavioral expression, even though the synapses are not the locus of information storage. \emph{In vitro} LTP typically decays to baseline within a few hours \citep[longer durations have been measured \emph{in vivo}, e.g.,][]{abraham02}. This implies that any behavioral memory expression that outlasts this time window is not mediated by storage at the synapse in question. I will discuss the issue of long-term storage further below. For now, the important point is that the evidence for synaptic storage of information is quite weak; in contrast, it is clear that synapses play an important role in memory expression.

\subsection{The trouble with synapses}

Although synapses are widely viewed as the site of memory storage, and synaptic plasticity is widely viewed as the basis of memory formation, the previous section has already indicated that decisive evidence for these views is lacking. In this section, I undertake a systematic critique of the synaptic model of memory.

\subsubsection{Timescales}

Hebbian rules for plasticity imply a temporal constraint: the presynaptic and postsynaptic neurons must fire close in time in order to produce a change in synaptic strength. How close? This question was answered by studies that measured LTP and LTD as a function of relative spike timing, finding that outside a window of about 40-60 milliseconds no plasticity occurs \citep{markram97,bi98}. This is known as \emph{spike-timing-dependent plasticity} (STDP).

This temporal constraint presents a serious problem, since animals can easily learn associations between stimuli separated by much longer intervals. The challenge then is to explain how synaptic plasticity can produce learning at behaviorally relevant timescales. One answer to this challenge involves neurons that generate sustained responses. If a stimulus can produce an ``eligiblity trace'' that outlasts the stimulus, then a conventional STDP rule can still produce an association \citep{drew06}. For example, calcium plateau potentials in apical dendrites could play this role \citep{bittner17}. Alternatively, a neuromodulator with slower kinetics, such as a dopamine, might provide the temporal bridge between stimuli \citep{izhikevich07,gerstner18}.

These mechanisms can extend the timescale of synaptic plasticity from milliseconds to seconds, but they cannot explain how learning is possible over longer timescales. Animals can acquire conditioned taste aversion when sickness follows ingestion by several hours \citep{smith67,revusky68}. In this case, one might argue that a long-lasting eligibility trace exists in the digestive system---the so-called ``aftertaste'' hypothesis. If true, this would allow a conventional STDP mechanism to explain taste aversion learning over very long delays. However, considerable evidence opposes the aftertaste hypothesis \citep[see][for a review]{rozin71}. Taste aversions can be established under physiological conditions in which a long-lasting trace is highly implausible. For example, only a single conditioning trial with a saccharin solution is needed to produce taste aversion, even when the delay to sickness is 12 hours, at which point the ingested saccharin has a negligible presence in the blood stream and digestive system. Aversion can also be learned to the high or low temperature of a drink, despite the fact that the fluid's temperature is quickly converted to body temperature after ingestion \citep{nachman70}. Another argument against the aftertaste hypothesis rests on the assumption that whatever the trace is, it must decay over time. Indeed, the strength of conditioned taste aversion declines with delay, consistent with other Pavlovian conditioning preparations. Presenting the conditioned stimulus again should presumably ``reactivate'' the trace, thereby rendering taste aversion stronger. In fact, the opposite happens: taste aversion is weakened, possibly due to a ``learned safety'' effect \citep{kalat73}.

This is not necessarily the end of the story, since an eligiblity trace could be maintained neurally even in the absence of a peripheral trace. For delays of up to 3 hours, conditioned taste aversion depends on sustained phosphorylation of calmodulin-dependent protein kinase II in the insular cortex \citep{adaikkan15}. The existence of such a trace, however, does not imply that conditioned taste aversion is mediated by STDP. It is unclear how such a mechanism could achieve the selectivity of learning observed in conditioned taste aversion (a point I discuss more in the next section). Another issue is that a very slowly decaying trace would have to terminate relatively abruptly after the aversive outcome in order to avoid producing LTD that would partially negate the antecedent LTP. Such precise (millisecond) timing for a trace that is assumed to be governed by kinetics operating on the timescale of hours presents a biophysical conundrum.

I have not yet confronted an even more fundamental timescale problem: \emph{there is no characteristic timescale for learning}. While it is true that associative learning is weaker when two stimuli are separated by a longer delay, the functionally relevant units of time are not absolute but relative. To understand this, consider the fact that spacing trials increases the rate of learning. Putting the delay-dependence of learning together with the spacing effect, we obtain the timescale invariance of learning: learning rate is typically constant across inter-stimulus delays as long as the inter-stimulus interval is proportionally rescaled \citep{gibbon77,ward12,gallistel00}. As pointed out by \citet{gallistel13}, synaptic plasticity does not in general exhibit this property. On the contrary, \citet{dejonge85} showed that the cumulative strength of LTP is weaker when the tetanic stimulations are separated by longer intervals. Other studies have observed a spacing effect in the same direction as behavior. \citet{zhou03} demonstrated stronger LTP with spaced stimulation intervals. However, unlike the spacing effect in behavioral studies of learning, the benefit of spacing diminished beyond about 5 minutes, such that the longest tested intervals (10 minutes) produced LTP comparable to that of massed presentation (0 minutes). Other studies demonstrating advantages for spaced stimulation also used the optimal (5 minute) interval without assessing longer intervals \citep{scharf02,woo03}.

\subsubsection{Selectivity}

One of the puzzles that any associative theory of learning must address is why associations appear to be selective. For example, in the conditioned taste aversion experiments using X-rays to induce sickness \citep[e.g.,][]{smith67,revusky68}, it takes about an hour for the the animals to feel sick. Everything the animal does in the intervening hour has closer temporal proximity to the sickness event, and should (according to the temporal logic of Hebbian association) form stronger aversions. Yet the animal learns to avoid the gustatory stimulus to which it was exposed rather than to any of these other stimuli \citep[for further discussion, see][]{seligman70,rozin71}.

In an elegant experiment, \citet{garcia66} showed that an audiovisual stimulus occurring at exactly the same time as the gustatory stimulus did not elicit avoidance. However, when an electric shock was used as the aversive outcome, animals avoided the audiovisual stimulus rather than the gustatory stimulus. This kind of selectivity does not by itself defeat all theories of learning based on synaptic plasticity, but it does impose some strong constraints. Synapses either have to be restricted to only those stimulus-selective neurons that are eligible for association, or the synapse has to somehow ``know'' which presynaptic inputs to ignore. Neither of these constraints has been systematically studied in the literature on synaptic plasticity.

\subsubsection{Content}

We have seen above how synaptic plasticity is sensitive to time. This temporal sensitivity is critically distinct from temporal coding: synapses do not provide a representation of time that is computationally accessible to downstream neurons \citep{gallistel13}. The distinction between sensitivity and coding matters because animals appear to use representations of time to guide their behavior, even in simple Pavlovian conditioning protocols \citep{savastano98,molet14}. A few examples will suffice to illustrate this fact.

\citet{barnet97} trained rats with either forward conditioning (cue1 $\rightarrow$ outcome) or backward conditioning (outcome $\rightarrow$ cue1). Consistent with previous studies, they found weaker responding to the cue in the backward condition.\footnote{Note that excitatory, albeit weaker, responding in backward conditioning is already problematic for an STDP model, which predicts only LTD.} In addition, they trained a second-order cue that preceded the first-order cue (cue2 $\rightarrow$ cue1), finding that responding to the second-order cue was \emph{stronger} in the backward condition. This puzzling observation makes sense under the hypothesis that animals had learned the temporal relationship between cue1 and the outcome, which they could then reuse to drive learning about cue2. Specifically, if the animals anticipate that the outcome will occur sooner following cue2 in the backward condition, they will exhibit a stronger response (consistent with the findings, already discussed, that shorter cue-outcome intervals produce stronger conditioning when the inter-trial interval is held fixed). This nicely demonstrates how an explicit representation of temporal expectation is made computationally accessible to another learning process. It is difficult to see how a purely associative account would predict the observed differences in response strength to the cues.

In a similar vein, \citet{cole95} used a trace conditioning procedure in which the outcome occurred 5 seconds after the offset of cue1. They then exposed animals to cue1 $\rightarrow$ cue2 pairings. The ensuing response to cue2 was \emph{stronger} than the response to cue1, despite the fact that cue1 had been directly paired with the outcome and cue2 had not. Again, this is deeply puzzling from an associative point of view, but makes sense if animals had learned a temporal expectation which rendered cue2 in closer proximity to the outcome compared to cue1.

Temporal coding exemplifies a more general point about the contents of memory: the brain cannot compute with information that it does not represent in memory \citep{gallistel11}. If we adopt the position that the brain learns some kind of content (e.g., temporal expectations), then this information must be stored in a format that is accessible to downstream computation. This logic applies to non-temporal representations. For example, pigeons can use representations of spatial relatinoships to identify food locations. If two landmarks (A and B) occur in a consistent spatial relationship, and then one of them (A) is placed in a particular spatial relationship to a food location, pigeons will search in the appropriate food location when exposed to landmark B, consistent with the idea that animals have learned a spatial map that supports integration of partial information \citep{blaisdell05,sawa05}. Similar results have been reported with rats \citep{chamizo06}. A parsimonious hypothesis is that these animals store the spatial coordinates of landmarks in an allocentric map.

The challenge for synaptic models of memory is to explain how information about content like time and space is stored in a format that supports the flexible behavior described above. One influential proposal is that these variables could be encoded in the activity state of neural populations. For example, elapsed time could be encoded by the stimulus-induced dynamics of population activity, and then decoded by downstream neurons \citep{karmarkar07}. Similarly, spatial position could be encoded by population activity in an attractor network \citep{burak09,samsonovich97}. Although these models make time and space computationally accessible, they don't solve the problem of long-term storage: how can information be retrieved long after persistent activity has subsided? Even over relatively short retention intervals (on the order of cellular time constants), information stored in persistent activity is quickly degraded due to noise \citep{burak12}. The same problem occurs for some models that encode spatial information using patterns of oscillatory interference \citep{zilli09}. Other models posit that very slowly varying activity states can retain long-term memories \citep{liu19,shankar12}. However, under some conditions memory can be retained even across periods in which stimulus coding cannot be detected in ongoing activity---so-called ``activity silent'' memory \citep{beukers21}.

A different line of theorizing employs synaptic plasticity in a central role for long-term storage of temporal and spatial information. I will focus on spatial information storage in the hippocampus, which has been extensively studied. The hippocampus contains neurons that selectively fire in particular spatial locations (``place cells''), and the spatial tuning of these cells is modifiable by synaptic plasticity \citep[see][for a review]{cobar17}. The question I seek to address here is how such a mechanism can store spatial memories in a durable format. It is known that the hippocampus is essential for the performance of memory-based navigation tasks such as the Morris water maze \citep{morris82,moser95}, including retention intervals on the order of weeks \citep{clark05}, making it plausible that the hippocampus is a long-term storage site for spatial information.

STDP can be used to learn a ``navigation map'' in which firing fields shift in the animal's traversal direction \citep{blum96,redish98}. The difference between coded and actual location defines a vector that approximately points towards the goal and is therefore useful for goal-directed navigation. An important challenge for this kind of model is that the map is quickly overwritten when the animal is exposed to multiple goal locations. \citet{gerstner97} solved this problem by allowing the receptive fields of hippocampal cells to be modulated by goal location. This solution presupposes that the enduring memories of goal location is stored outside the hippocampus; the hippocampal neurons are, in effect, a read-out of these memories. Gerstner and Abbott do not provide an account of how these memories are stored.

\citet{hasselmo09} has suggested that route memory can be linked to sensory cues by synaptic plasticity between sensory representations and place cells. A basic problem for this kind of model is that goal-directed navigation can operate in the absence of sensory cues: rats can navigate complex mazes even when they are deprived of sight \citep{carr17b} and smell \citep{carr17c}, provided that they have been previously familiarized with the mazes. Place cells likewise maintain their firing fields in a dark environment when rats were previously provided with visual experience of the environment \citep{quirk90}. Moreover, place cells which were initially under sensory control by a distal landmark (e.g., cells with firing fields that rotate in correspondence with a cue outside the arena) will continue to fire in their preferred location even when the distal landmark has been removed completely \citep{muller87,okeefe87,shapiro97}.

In summary, the problem of content is the problem of storing the relevant content in a computationally accessible format. Existing models of memory based on synaptic plasticity have not yet offered a comprehensive account of how spatial and temporal content could be stored and retrieved in the service of flexible navigation.

\subsubsection{Memory persistence and synaptic instability}

As mentioned earlier, LTP typically decays to baseline on the order of hours. Yet some childhood memories last nearly our entire lifetime. How can we resolve this discrepancy?

Before discussing possible resolutions, let us dive a bit deeper into the nature of the problem: why does LTP decay? Synapses are typically contained in small dendritic protrusions, called \emph{spines}, which grow after induction of LTP \citep{engert99}. Spine sizes are in constant flux \citep[see][for a review]{mongillo17}. Over the course of 3 weeks, most dendritic spines in auditory cortex will grow or shrink by a factor of 2 or more \citep{loewenstein11}. In barrel cortex, spine size changes are smaller but still substantial \citep{zuo05}. Spines are also constantly being eliminated and replaced, to the extent that most spines in auditory cortex are replaced entirely over the same period of time \citep{loewenstein15}. In the hippocampus, the lifetime of spines is even shorter---approximately 1-2 weeks \citep{attardo15}.\footnote{Some of the discrepancies between estimates of spine turnover rates may reflect methodological choices, such as the type of cranial window \citep[see][]{xu07}.} Importantly, much of the variance in dendritic spine structure is independent of plasticity pathways and ongoing neural activity \citep{minerbi09,dvorkin16,quinn19,yasumatsu08}, indicating that these fluctuations are likely not generated by the covert operation of classical plasticity mechanisms. Collectively, these observations paint a picture of profound synaptic instability.

Another set of clues comes from studies of animals whose brains undergo radical remodeling during certain phases of their life cycle \citep{blackiston15}. Holometabolous insects undergo complete metamorphosis from the larval to adult stages, including large-scale cell death, neurogenesis, and synaptic pruning. For example, olfactory projection neurons in the antennal lobe of the fruit fly \emph{Drosophila} persist from the larval to the adult stage, but their synapses with mushroom body neurons are disassembled during metamorphosis \citep{marin05}. Furthermore, the mushroom body itself is substantially reorganized during metamorphisis \citep{armstrong98}. Nonetheless, conditioned olfactory avoidance acquired during the larval stage is retained in the adult stage \citep{tully94}. This is made more puzzling from a synaptic perspective by the fact that disrupting the inputs to the mushroom body impairs the acquisition olfactory conditioning \citep{heisenberg85}. Apparently these synapses are required for learning but not for long-term storage.

Some mammals undergo significant brain remodeling during hibernation. Perhaps the most well-studied example is the family of European and arctic ground squirrels, whose brains experience massive dendritic spine retraction during hibernation \citep{popov92,ohe06}, resulting in a 50-65\% loss of synapses \citep{ohe07}. Remarkably, this pruning is reversed within 2 hours of arousal. While studies of memory retention across hibernation have yielded mixed results \citep[see][]{blackiston15}, evidence suggests that at least some information learned prior to hibernation is retained by ground squirrels after arousal \citep{millesi01}. Again, the challenge raised by such results is how behaviorally expressed memories could be more persistent than their putative synaptic storage site.

Even more dramatically, memories can persist in some species across decapitation and regeneration. Planarians (a class of flatworms) are distinguished by their incredible powers of regeneration from tiny tissue fragments \citep[in some cases less than 0.4\% of their body mass;][]{morgan01}. Planarians are also capable of learning: if repeatedly shocked after presentation of a light, a planarian will learn to avoid the light \citep{thompson55}. Now suppose you cut off a planarian's head after it has learned to avoid light. Within a week, the head will have regrown. The critical question is: will the new head remember to avoid light? Remarkably, a number of experiments, using light-shock conditioning and other learning tasks, suggested (albeit controversially) that the answer is yes \citep{mcconnell59,corning61,shomrat13}. What kind of memory storage mechanism can withstand utter destruction of brain tissue?

Now that we understand the challenges posed by synaptic instability, we can discuss some possible solutions. The review that follows is not exhaustive; for present purposes, it suffices to briefly introduce the key ideas and supporting evdience.

A standard answer appeals to \emph{systems consolidation}, the process by which memories are gradually transferred from a temporary store in the hippocampus to a more durable representation in the neocortex \citep{squire15}. This kind of solution still begs the question: how are the new representations durable if they too depend on transient synapses? One possibility is that distributed cortical memories are more robust to degradation compared to hippocampal memories. Another possibility is that multiple traces co-exist, perhaps even in the hippocampus itself \citep{nadel00}, which would also confer robustness.

A number of modeling studies have explored synaptic learning rules that attempt to overcome the problem of instability. One approach relies on dendritic spines with different lifetimes. Experiments have shown that some spines are stable on the timescale of months \citep{yang09}. Models with multiple timescales of plasticity could potentially harness this property to construct a form of synaptic consolidation, whereby memories are passed through a cascade of progressively more stable storage sites \citep{fusi05,benna16}.

Multiple timescales can also be realized at the molecular level using bistable switches \citep{crick84}. The most well-known example of a hypothetical switch is calmodulin-dependent protein kinase II (CaMKII), which assembles into a holoenzyme structure consisting of 12 CaMKII subunits \citep{lisman12}. LTP induction leads to phosphorylation of several subunits, and this phosphorylation state is maintained through autophosphorylation by neighboring subunits. In this way, the switch can persistently maintain memories. In support of this hypothesis, inhibiting CaMKII can reverse LTP \citep{barcomb16,sanhueza11}. At the behavioral level, inhibiting CaMKII in the hippocampus after learning impairs conditioned place preference \citep{rossetti17}, and inhibiting CaMKII in the nucleus accumbens impairs amphetamine sensitization \citep{loweth13}.\footnote{It should be noted that, while a role for CaMKII in LTP induction is widely supported, some studies find that it is not necessary for memory maintenance/expression \citep{malinow89,murakoshi17,chen01}.}

Another class of models makes use of the fact that connected neurons typically have multiple synaptic contacts \citep{fares09}. With appropriate coordination, these synapses can collectively store memories that outlive the lifetime of individual synapses \citep{deger12,fauth15}. Alternatively, clusters of receptors on a single synapse can be coordinated (by correlation of receptor insertation rate between neighboring locations) to extend memory lifetime \citep{shouval05}.

Some models achieve stabilization using memory reactivation (``rehearsal''), which can effectively repair degraded synapses \citep{wittenberg02,acker19,shaham21,fauth19}. Under some conditions, the correlation structure of reactivated memories can even strengthen memories that aren't reactivated \citep[so-called ``implicit rehearsal'';][]{wei14}.

Finally, memories could be maintained at the network level if some properties of the network remain invariant to changes in individual synapses \citep{susman19} or are protected by some form of compensatory plasticity \citep{raman21}. \citet{susman19} studied how the complex eigenvalues of the network dynamics are affected by homeostatic plasticity and synaptic noise. They showed that the complex part of the eigenvalue spectrum, which determines the structure of fluctuations around the set-points of network activity, encodes long-term memories that are robust to erasure by homeostasis. They then showed that STDP mainly affects the imaginary part of the spectrum. These memories are defined not by fixed points (as in classical models of memory, such as the Hopfield network), but by limit cycles---time-varying stable states.

In summary, ``garden variety'' synaptic instability (i.e., the kind that appears under typical conditions of brain function) appears to be a surmountable challenge for computational models of memory persistence. The more radical varieties of synaptic stability reviewed above (e.g., during metamorphosis, hibernation, and decapitation) are not as easily dealt with. They suggest a memory storage site that transcends synapses---and possibly even cells---entirely.

\subsubsection{Savings}

Many forms of learning show ``savings'' after the original behavioral expression of learning has disappeared. For example, a conditioned response can be exinguished by repeatedly presenting the conditioned stimulus alone; when the stimulus is then reconditioned, acquisition is faster \citep{napier92,ricker96}, demonstrating that the memory has not been entirely lost due to extinction. Rapid reacquisition has been observed in many different preparations, ranging from habituation of the galvanic skin response \citep{davis34} to instrumental learning \citep{bullock53} and visuomotor adaptation \citep{krakauer05}.

Savings also appears in other ways \citep[see][for a review]{bouton04}. \citet{pavlov27} noted that extinguished responses spontaneously recover after extinction, a finding that has been demonstrated repeatedly \citep{rescorla04}. Like rapid reacquisition, spontaneous recovery has also been observed for habituation \citep{prosser36}, instrumental learning \citep{rescorla97}, and visuomotor adaptation \citep{kojima04}.

A natural synaptic model of extinction assumes that the relevant synapse is depotentiated---i.e., extinction reverses the synaptic effects of conditioning. This explanation is consistent with the aforementioned endocytosis of AMPA receptors during extinction \citep{kim07,clem10}. If true, then we would not expect any savings, because the associative memory is erased. Indeed, LTP does not exhibit savings: reinduction after depotentiation proceeds at the same pace as original induction \citep{dejonge85}. Clearly, this property of LTP is at variance with the behavioral findings, motivating alternative hypothesess about the nature of extinction, according to which extinction memories are stored separately rather than erasing the original memory \citep[e.g.,][]{milad12,gershman17b}. These alternative hypotheses do not, without further elaboration, explain how memory is stored at the level of cells.

Studies of experimental amnesia provide another source of evidence against a synaptic model of savings. It is commonly believed that long-lasting (``late'') LTP requires protein synthesis, since inhibiting protein synthesis prevents LTP induction \citep{stanton84}. Accordingly, memory is almost completely obliterated within a few hours following training under systemic protein synthesis inhibition \citep{squire72}. However the same study demonstrated that spontaneous recovery occurs when animals are tested 3 days later. Evidently, blocking LTP was not sufficient to eliminate the memory permanently \citep[see also][]{power06,lattal04}. Strikingly, the protein synthesis inhibitor can itself act as a retrieval cue, stimulating retrieval of memories that were supposedly erased by the same treatment \citep{bradley88,briggs13,gisquet15}. The next section discusses other cases of recovery from experimental amnesia.

\subsubsection{Dissociations between memory and synaptic integrity}

If synapses are the storage site of memory, then memories should not survive the disruption of synapses. Several reports challenge this assumption.

Sensitization of siphon withdrawal in the sea slug \emph{Aplysia} is produced by long-term facilitation (LTF) at synapses between sensory and motor neurons \citep{frost85}. LTF can also be induced in dissociated cell cultures using pulses of serotonin, which mediates sensitization, and is accompanied by growth of new presynaptic varicosities \citep{glanzman90}. LTF can be reversed by inhibiting protein synthesis following a brief `reminder' pulse of serotonin \citep{cai12}, which causes the reversion of presynaptic varicosities to their pretraining number \citep{chen14}. However, the reverted synapse differs morphologically from the pretraining synapse, suggesting that protein synthesis inhibition does not simply undo the effects of LTF. Even more remarkably, Chen and colleagues showed that sensitization memory can survive synaptic degradation: a small amount of additional training following the administration of protein synthesis inhibition was sufficient to renew sensitization in previously trained animals, but not in naive animals. These findings \citep[echoing others in the rodent literature;][]{power06,lattal04} call into question the widely held belief that the sensory-motor synapse is the storage site of sensitization memory in \emph{Aplysia} \citep{kandel01}.

Protein synthesis inhibitors also disrupt contextual fear conditioning in rodents when administered immediately after training. When injected locally into the hippocampus, they specifically suppress LTP at postsynaptic neurons that were active during memory formation \citep[so-called ``engram cells'';][]{ryan15}. Surprisingly, optogenetic reactivation of these engram cells can produce conditioned fear even after protein synthesis inhibition. In other words, engram reactivation can induce recovery from amnesia. This phenomenon was not specific to the hippocampus; Ryan and colleagues also observed that conditioned fear to a tone could be optogenetically reactivated after local injection of protein synthesis inhibitors into the lateral amygdala.

It is uncontroversial that protein synthesis inhibitors disrupt synaptic plasticity, and that this disruption interferes (at least temporarily) with memory expression. What these studies demonstrate is that the link between synaptic plasticity and memory \emph{storage} is more tenuous. A number of authors have pointed out that commonly used protein synthesis inhibitors such as cycloheximide and anisomycin have other cellular effects, including apoptosis, inhibition of catecholamine function, control of post-translational protein modification, and even the \emph{elevation} of protein synthesis due to negative feedback regulation \citep{routtenberg05,rudy06}. Furthermore, a large (and largely forgotten) literature shows that the amnestic effects of cycloheximide and anisomycin can be attenuated by a wide range of treatments (e.g., amphetamines, caffeine, nicotine) that do not reverse the inhibition of protein synthesis \citep{martinez81}. We should therefore not accept uncritically the thesis that protein synthesis inhibitors achieve their amnestic effects through destabilization of a memory trace.

Beyond protein synthesis inhibitors, \citet{shors97} have noted that ``among the dozens of compounds shown to retard the induction of LTP in the hippocampus, only a few directly influence learning, and many others have no effect on learning'' (p. 606). A similar conclusion is suggested by genetic deletion studies. For example, hippocampal LTP is debilitated in mice lacking the $\gamma$ isoform of protein kinase C \citep{abeliovich93b}, yet these mice exhibit relatively normal learning of the Morris water maze, a hippocampus-dependent task \citep{abeliovich93}. Hippocampal LTP is also debilitated in mice lacking the GluR-A submunit of the AMPA receptor, yet spatial learning in the water maze was unimpaired \citep{zamanillo99}. The same story has been reported for brain-derived neurotophin factor knockout mice \citep{montkowski97}. Mice lacking endothelial nitric oxide synthase actually show \emph{superior} performance in the Morris water maze \citep{frisch00} despite deficient LTP \citep{wilson99}. Taken together, these results show that synaptic plasticity and memory performance can be dissociated in many different ways.

\subsubsection{Intrinsic plasticity: from excitability to coding}

It has long been known that non-synaptic (or \emph{intrinsic}) plasticity occurs in neurons \citep{mozzachiodi10,titley17}. Most studies have focused on plasticity of neuronal excitability, which can be meaured in different ways (most commonly the threshold or slope of the neuron's input-output function). In one early demonstration, \citet{woody73} measured the amount of current required to elicit a spike in cortical motor neurons after eyeblink conditioning. The current threshold was reduced for cells projecting to eyeblink musculature.

Changes in intrinsic excitability are often coupled with changes in synaptic strength. For example, sensitization of siphon withdrawal in \emph{Aplysia} is accompanied by both facilitation of sensorimotor synapses (reviewed in the previous section) and increases in sensory neuron excitability \citep{cleary98}. Increased excitability can augment synaptic strength by broadening presynaptic action potentials, resulting in more neurotransmitter release \citep{byrne96}. Excitability may also control the threshold for induction of synaptic plasticity \citep{triesch07}, or serve as an elgibility trace to synaptic plasticity across long delays between presynaptic and postsynaptic firing \citep{janowitz06}. On the other hand, changes in excitability and synaptic strength can sometimes be dissociated \citep[see][for a review]{mozzachiodi10}, suggesting that they may play distinct functional roles.

Intrinsic plasticity extends beyond changes in excitability, as illustrated by studies of Purkinje cells in the cerebellum. These cells tonically inhibit the cerebellar output nuclei; pauses in firing produce overt movement, including eyeblinks. Animals can learn to produce eyeblinks, via pauses in Purkinje cell firing, that anticipate an aversive airpuff delivered to the eye when it is preceded by a reliable stimulus cue. According to standard models of cerebellar conitioning \citep{marr69,albus71,bullock94}, granule cells are thought to convey a temporal pattern of activity (e.g., neurons with tuning to different time intervals) that can be used to predict airpuff timing. The airpuff signal is conveyed to Purkinje cells by climbing fibers from the inferior olivary nucleus. Granule cell synapses that are active prior to climbing fiber spikes undergo LTD, such that subsequent granule cell activity inhibits the Purkinje cell.

For a number of reasons \citep[see][for a recent review]{johansson19}, this model is unlikely to be correct. One reason is that pharmacological or genetic inhibition of LTD appears to have little effect on adaptive timing of conditioned eyeblinks \citep{welsh05,schonewille11}, and in some preparations LTD does not even occur \citep{johansson18}. Even when LTD does occur, its properties are mismatched to the properties of Purkinje cell firing: LTD emerges on the order of minutes, whereas conditioned responding of Purkinje cells (similar to the conditioned eyeblink) emerges on the order of hours \citep{chen95,jirenhed07}. Another reason is that Purkinje cells appear to store information about the timing of input signals in a cell-intrinsic format \citep{johansson14}. I now explain this experiment in more detail.

Johansson and colleagues directly stimulated the axons of granule cells (the parallel fibers) and the climbing fibers while measuring Purkinje cell activity in decerebrate ferrets. The same parallel fibers were stimulated every 10 milliseconds for a 200-800 millisecond interval prior to climbing fiber stimulation. This feature of their experiment is critical, because it means that there is no temporal pattern encoded by the granule cell population.\footnote{Further evidence against a granule cell time code is provided by studies showing that stimulation of inputs to granule cells tends to produce a short-latency respone with little variation across cells \citep{jorntell06}.} Any LTD that might be happening at parallel fiber synapses would be expected to affect all the stimulated synapses in a uniform way, rendering the learning mechanism useless for predicting the timing of the climbing fiber input. Nonetheless, Purkinje cells in this preparation acquired exquisitely timed anticipatory responses to parallel fiber input, and can even acquire information about multiple time intervals \citep{jirenhed17}. These findings imply that the learned timing information is intrinsic to the cell.

\subsubsection{Memory transfer}

Few topics in neuroscience are more bizarre and controversial than memory transfer. The history is too long and complex to fully recount here \citep[see][]{travis81,setlow97}, but I will try to extract some lessons and connect this history to more recent developments.

As discussed earlier, McConnell and his collaborators showed that planarians can retain conditioned fear of a light stimulus across decapitation \citep{mcconnell59}. At the time, neurobiologists were not yet locked into synaptic theories (LTP had not yet been discovered). Much attention was focused on the hypothesis (discussed further below) that the storage site was biochemical, possibly RNA or some other macromolecule. This seemed appealing as an account for the planarian data, since the hypothetical memory molecule could potentially survive decapitation by being distributed throughout the nervous system \citep{mcconnell68}. In particular, McConnell speculated that memories were stored inside neoblasts---undifferentiated, pluripotent cells that circulate throughout the body of the worm, are rich in RNA, and provide the basis of its astounding regenerative powers. The biochemical hypothesis also seemed to offer a way of addressing Lashley's well-known failure to anatomically localize the memory trace based on lesion data \citep{lashley29}.

Inspired by the biochemical hypothesis, \citet{mcconnell62} took advantage of naturally occurring cannibalism between planarians, feeding trained subjects to untrained subjects, and finding that the untrained subjects exhibited conditioned fear of the light stimulus on the first day of training. The untrained subjects had apparently ingested a memory!

Around the same time, \citet{corning61} pursued the RNA hypothesis more directly by bathing decapitated planarians in ribonuclease (an enzyme that hydrolizes RNA). Under this treatment, previously trained planarians did not retain memory after decapitation. A few years later, it was shown that directly injecting RNA extracted from trained planarians into untrained planarians could produce a transfer effect \citep{jacobson66}. The study also included a pseudo-conditioning control group, which received the same number of light and shock stimuli but without systematic temporal contiguity between their occurrences. Pseudo-conditioned RNA donors did not produce a transfer effect in recipients, supporting the conclusion that the memory being transferred encoded specific information about the relationship between stimuli. This control was important for ruling out the alternative hypothesis that transfer effects were explained entirely by sensitization to the light or shock \citep{walker66,hartry64}. It is also important to note that the transfer effect was observed without any additional training, which argues against a metaplasticity interpretation (acceleration of new learning without retention of old learning).

Despite these compelling demonstrations of memory transfer, McConnell's research was eventually dismissed as a failure \citep{rilling96}. In a recent review, \citet{josselyn17} reiterate this assessment:
\begin{quote}
    Many attempted to replicate McConnell's memory transfer findings. Some replicated his findings, but many did not. McConnell’s work was widely criticized for being poorly designed and poorly controlled. (p. 4653)
\end{quote}
The scholarly record supporting this assessment is quite thin. Only a small handful of published studies attempted to replicate the planarian memory transfer effects, and of these only a single one \citep{walker66b} failed to replicate the result. As already noted, the \citet{jacobson66} study effectively controlled for several potential confounds, yet it is rarely mentioned in the literature on memory transfer.\footnote{Despite being published in \emph{Nature}, the Jacobson paper was omitted from several contemporary surveys of the memory transfer literature \citep{mcgaugh67,smith74}, as well as more recent surveys \citep{morange06,colacco18}.}

There are several reasons why the planarian memory transfer work was unfairly maligned. First, McConnell had a penchant for self-publishing his findings in his own satirical journal, \emph{Worm Runner's Digest} (although the first transfer study was eventually published in a peer-reviewed journal). Second, there were disputes about whether planarians could acquire classical conditioning at all, let alone transfer this learning, but these were addressed by the introduction of appropriate control conditions \citep{block67,jacobson67}. Third, it appears that McConnell and other planarian researchers were victims of guilt by association with later studies of memory transfer, particularly with rodents, that failed to replicate reliably. I now turn to these studies.

The first reports of memory transfer between rodents were reported by four separate laboratories in 1965 \citep{babich65,babich65b,fjerdingstad65,reinis65,ungar65}. By the mid-1970s, hundreds of rodent memory transfer studies had been carried out. According to one tally \citep{dyal71}, approximately half of the studies found a positive transfer effect. This unreliability, combined with uncertainty about the underlying molecular substrate and issues with inadeqaute experimental controls, were enough to discredit the whole line of research in the eyes of the neuroscience community \citep{setlow97}.

There are a number of important caveats to this conclusion. One is that many of the failed replications transferred RNA extracts based on the hypothesis that RNA was the memory molecule. \citet{ungar74}, in contrast, hypothesized that the memory molecule mediating transfer was a peptide (a short amino acid sequence). Consistent with this hypothesis, Ungar's original report of memory transfer \citep{ungar65} showed that incubating brain homogenates with chymotrypsin (which hydrolizes peptide bonds) prior to transfer eliminated the transfer effect, whereas incubating with ribonuclease did not.

Another problem for the RNA hypothesis is that most RNA cannot cross the blood-brain barrier. Accordingly, intra-peritoneal injections of RNA extracts did not produce a measurable change in brain RNA levels \citep{luttges66,guttman72}. Ungar believed that his peptide hypothesis did not suffer from this problem, arguing that peptides are typically small enough to cross the blood-brain barrier. However, it is now understood that the blood-brain barrier permeability to peptides is low, and that brain capillaries have high levels of peptidases \citep{pardridge83}.

If correct, the peptide hypothesis (or indeed any non-RNA hypothesis) would render the replication failures using RNA extracts irrelevant. \citet{ungar70} suggested that the memory-encoding peptide or protein might be bound in a complex to RNA, separated only after injection, which would explain how RNA could be incidentally associated with transfer effects despite not conveying learned information.

A second caveat is that the sheer variety of transferred behaviors---including habituation, passive dark avoidance, active footshock avoidance, taste aversion, conditioned magazine approach, bar pressing, wire climbing, spatial alternation, spatial discrimination, size discrimination, and color discrimination---and the variety of species tested---including worms, hamsters, mice, rats, goldfish, and even chickens---makes it difficult to completely dismiss the transfer phenomenon as a peculiarity of some particular preparation. The variability of preparations also provides insight into the conditions for successful transfer. \citet{smith74} reviewed a wide range of procedural variables and concluded that many made little difference to the outcome or were equivocally related to the outcome. The most important variables related to the brain extracts, such as extract dosage, chemical composition, heat, purification procedure, among others. Because these variables differed considerably across laboratories, it is not straightforward to discriminate between replication failures and novel experimental findings.

After a long hiatus, memory transfer work has recently made a comeback, using new tools from molecular biology and new model systems. \citet{bedecarrats18} showed that sensitization in \emph{Aplysia} could be transferred to untrained subjects by injection of RNA. This study went beyond the earlier literature in several ways. One is that it identified a neurophysiological correlate of transfer: sensory neurons (but not motor neurons) of RNA recipients exhibited increased excitability. RNA injection also increased the strength of a subset of sensorimotor synapses, although this increase was not significant at the population level.

Another recent study showed that learned avoidance behavior in the nematode \emph{C. elegans} can be transferred \citep{moore21}. Naive animals are initially attracted to media containing the toxic bacteria \emph{P. aeruginosa}, but they learn to avoid it after exposure. Moore and colleagues showed that ingestion of homogenates from trained animals produced avoidance behavior in naive animals. Transfer of avoidance behavior is plausibly a naturally occurring phenomenon, because exposure to \emph{P. aeruginosa} can result in autophagy and ultimately autolysis. The lysates may then be consumed by other animals. Another remarkable feature of this model system is that the mechanism of transfer has been identified: pathogen avoidance is encoded by a small RNA (P11) packaged into virus-like particles that enable extracellular transmission. I will discuss this mechanism further when I consider the RNA hypothesis in more detail.

Memories can be transferred not only ``horizontally'' between individuals of the same generation, but also ``vertically'' between individuals of different generations \citep[see][for a review]{miska21}. Horizontal and vertical forms of transfer appear to rely on some of the same intracellular mechanisms. For example, the P11 RNA mediating horizontal transfer of pathogen avoidance in \emph{C. elegans} also mediates vertical transfer \citep{kaletsky20}; indeed, \citet{moore21} showed that horizontally transferred memories are then vertically transferred, and both depend on the \emph{Cer1} retrotransposon that encodes the virus-like particles encapsulating P11. This raises the intriguing possibility that there exists a common code for memory storage that supports both forms. Conceivably, vertical transfer is the evolutionarily more ancient form (used even by prokaryotic organisms like bacteria, as well as by plants and fungi) which was co-opted by more complex organisms to support horizontal transfer. This would explain why long-term memory storage seems to rely on epigenetic mechanisms like DNA methylation and histone modification, which play central roles in transgenerational inheritance \citep{heard14}. I discuss these mechanisms further below.

For our purposes, the important implication of memory transfer research is that it should not be possible if memories are stored synaptically, since synapses cannot be transferred. Even if one has reservations about some of the early transfer research, it seems difficult to argue that none of the hundreds of results (including from more technically sophisticated recent experiments) are real.

\subsubsection{Learning in single cells}

If single cells can learn, then they must be using a non-synaptic learning mechanism. Like memory transfer, the question of single cell learning has a long and controversial history, recently surveyed in \citet{gershman21}. The conclusion of that survey was that single cells likely do learn.

Some of the earliest studies were carried out by the zoologist Herbert Spencer Jennings and reported in his seminal book, \emph{Behavior of the Lower Organisms} \citep{jennings06}. Jennings examined a wide range of behaviors in various simple organisms, including unicellular ciliates like \emph{Stentor} and \emph{Paramecium}. Jennings \citep[see also][]{dexter19} demonstrated that when repeatedly stimulated, \emph{Stentor} exhibited a regular sequence of distinct avoidance behaviors. This suggests that the response to stimulation was memory-based: the same external stimulus produced different behavioral output as a function of stimulation history.

In the half century that followed Jennings' work, there were sporadic attempts to study learning in \emph{Paramecium}, with mixed results. The most systematic program of research was undertaken by Beatrice Gelber, a neglected pioneer highlighted in our survey \citep{gershman21}. Gelber developed an appetitive conditioning paradigm in which a platinum wire was repeatedly swabbed with a bacterial suspension (a food source) and then dipped into a culture containing \emph{Paramecia}. Gelber found that cells would attach to the wire after training, even when it had not been swabbed \citep{gelber52}. They would not produce this behavior at the beginning of training or when only exposed to the wire by itself. She interpreted these results as indicating that the cells had acquired a conditioned response to a neutral stimulus (the wire).

In a series of subsequent experiments, Gelber mapped out the effects of several quantitative parameters, such as the spacing of trials \citep{gelber62} and the retention interval prior to the wire-only test \citep{gelber58}. Her research was criticized for suffering from confounds, such as the possibility that the \emph{Paramecia} or bacteria left a chemical residue during training that could influence subsequent behavior on the test trial \citep{jensen57,katz58}. She went to great lengths to address these confounds \citep{gelber57}, but ultimately this was not enough to convince skeptics. 

\citet{hennessey79} repeatedly paired a vibrational stimulus with an alternating current and showed that Paramecia could be trained to produce an avoidance response to the vibration. Importantly, they also showed that this conditioning effect could not be explained as an artifact of sensitization or pseudo-conditioning. Their preparation had the advantage over Gelber's in that the electrical field (unlike the bacteria) did not produce any chemical byproducts. Finally, they showed that the conditioned response was retained for 24 hours, confirming that the cells had acquired a long-term memory.

So far I have been discussing data from unicellular organisms, but learning has also been investigated in single cells dissociated from multicellular organisms. The biochemist Daniel Koshland followed this approach using the rat adrenal cell line known as PCI2 \citep[reviewed in][]{morimoto91}. Although these cells are not neurons, they share an embryological origin with neurons (in the neural crest), and have been used as a model of neurosecretion due to the fact that they carry vesicles of catecholamines and acetylcholine. Neurosecretion habituates after repetitive depolarization \citep{mcfadden90} or cholinergic stimulation \citep{mcfadden90b}. After several hours without stimulation, the secretory response rebounds, but only partially; this indicates that the cell has maintained a long-term memory trace of the stimulation history. Amazingly, a single cell can maintain multiple memory traces, as evidenced by the fact that habituation to cholinergic stimulation has no effect on habituation to depolarization, and vice versa.

While we still do not understand fully the learning capabilities of single cells, the totality of evidence makes a strong case that learning does occur in single cells. The mechanisms underlying such learning are still poorly understood. I now turn to a discussion of possible mechanisms.

\section{In search of the memory molecule}

If the preceding sections convinced you that molecular mechanisms are at play in memory storage, the next task is to idenfity them. I will focus on a subset of the possible mechanisms that have been most extensively studied, omitting some (e.g., peptides and microtubules) that are potentially important but have received less direct support.

\subsection{RNA and other cytoplasmic macromolecules}

The emergence of modern molecular biology in the 1940s, culminating in Watson and Crick's discovery of the DNA double helix in 1953, led to a rejuvenation of ideas about the nature of the memory trace. Even before Watson and Crick's discovery, it was understood that cells have biochemical mechanisms for information storage. \citet{gerard53} drew an analogy to immune cells, which store information about pathogens long after antibody levels have subsided. \citet{katz50} suggested that biochemical memory traces in the nervous system might consist of nuclear proteins, modified as a consequence of experience. Other authors suggested cytoplasmic macromolecules as memory traces \citep[see][for a review of these early proposals]{uphouse74}.

The impetus for macromolecular theories, and the RNA hypothesis in particular, came from several directions. One was the finding, reviewed earlier, that degrading RNA with ribonuclease impaired the retention of memory in decapitated planarians \citep{corning61}. Another was the rather bizarre finding that administration of yeast RNA (as well as DNA) to elderly patients seemed to improve their memory \citep{cameron58}; this finding was reinforced by subsequent studies in rodents \citep{cook63,solyom68}. While these results may have reflected a kind of general enhancement of cognitive or physical activity rather than specific enhancement of memory \citep[see][]{corson66,wagner66}, the contemporaneous memory transfer results reviewed earlier suggest a possibly more specific role for RNA in memory. Finally, changes to RNA were observed following learning \citep{hyden62,hyden64}, and inhibition of RNA impaired learning \citep{dingman61,chamberlain63}.

Broadly speaking, macromolecular theories fell into two classes: switchboard and direct coding theories. Switchboard theories viewed macromolecules as playing the role of a switch or connector, guiding the flow of information between neurons \citep{szilard64,rosenblatt67,ungar68,ungar70}. These theories align more closely with modern views of synaptic memory, according to which macromolecules play a supportive role (e.g., in determining synaptic strength). A basic difficulty facing such theories is explaining the memory transfer results. For example, \citet{ungar68} argued that the hypothetical molecular connectors provide a ``trail'' through genetically determined neural pathways:
\begin{quote}
    The situation could be compared to that of a traveler who is given a jumble of paper slips identifying the different intersections at which he has to turn to reach his destination. With the help of a map which shows the location of the intersections designated by the identification marks, he could easily find the correct sequence in which he should take them to reach his goal. For the learning process, the map is the genetically determined neural organization. When the connectors included in the extract injected into the recipient animal fall in place in the neurons homologous to their sites of origin, the nerve impulse follows the trail created by the training of the donors. (p. 229)
\end{quote}
Ungar's account rested largely on the principle of chemospecificity \citep{sperry63}, according to which neurons use a chemical labeling system to recognize which pathnways to form during development. When neurons are coactivated, they exchange chemical signals (what Szilard dubbed ``transprinting'') that allows them to form a connector representing the memory. However, Ungar did not explain precisely how the molecular connectors ``know'' where to go after injection, or how they would reprogram target cells.

Direct coding theories posited that macromolecules functioned as the terminal storage site for information, akin to how DNA is the terminal storage site for genetic information. Direct coding theorists appreciated that RNA likely played some role in storage, but it was unclear whether RNA itself was the terminal storage site. At the time these theories were first proposed (the early 1960s), the understanding of RNA was restricted mainly to its role in protein synthesis (i.e., messenger RNA). The first non-coding RNA was not discovered until the mid-1960s, and it would be several decades before the diversity of non-coding RNA was appreciated. Thus, apart from early speculations about coding via RNA \citep[e.g.,][]{gaito61,gaito63}, direct coding theories mainly focused on protein products as the terminal storage sites \citep{hyden61,briggs62,landauer64}. On this view, RNA plays an intermediate role in the construction of durable macromolecular codes. This point of view is buttressed by modern data on post-translational protein modification, which suggests that the combinatorial space of protein ``mod-forms'' provides a massive storage capacity for cellular history \citep{prabakaran12}.

One of the most intriguing direct coding theories was proposed by \citet{landauer64}, who conceived of neurons as electronic filters tuned to particular input frequencies. If the frequency tuning of a neuron is determined by some biochemical signature (Landauer was vague on this point), then changes in RNA composition would generate (via protein synthesis) changes in the neuron's signature and hence its frequency tuning. Landauer suggested that during periods of spiking activity, a neuron's membrane becomes permeable to RNA molecules originating in glial cells. If the spiking activity of the neuron could somehow select RNA that produce particular signatures, then the neuron could acquire long-term frequency tuning. Landauer's ideas, which were ahead of his time and remain neglected, appear to presage more recent theories of biolectric memory codes \citep{fields18}.

The direct coding version of the RNA hypothesis is attractive on theoretical grounds due to the fact that polynucleotide sequences are energetically cheap to modify, can be read by intracellular processes at fast timescales, and \citep[depending on secondary structure;][]{nowakowski97} can be thermodynamically stable \citep{gallistel17}. The addition of a single nucleotide to a polynucleotide sequence costs 1 ATP and can buy 2 bits of information. By contrast, synaptic modification costs $\sim4$ orders of magnitude more ATP \citep{karbowski19}, but by one estimate can only buy around 4-5 bits of information per synapse \citep{bartol15}.

Arguments against the RNA hypothesis appeared almost immediately after it was first proposed. \citet{briggs62} pointed out that the experimental evidence was largely indirect and could not rule out the plausible alternative hypothesis that RNA mediated memory effects via its conventional role in protein synthesis \citep[see also][]{dingman64}. They also argued that this protein synthesis pathway offered few opportunities for systematic RNA modification in the service of memory storage. Their conclusion was based mainly on the understanding of messenger RNA available at the time; as already noted, the diversity of RNA and the abundance of non-coding RNA was not yet appreciated. Other arguments against RNA as a memory molecule included its inability to cross the blood-brain barrier (raising questions about the RNA memory transfer experiments, as mentioned earlier), its susceptibility to degradation by endogenous ribonucleases, and the lack of experimental evidence for functional specificity.

These arguments were sufficient to kill off the RNA hypothesis for the next half-century. However, recent developments have brought it back to life. One development, already mentioned, was the finding that RNA is responsible for transfer of pathogen avoidance between nematodes \citep{moore21}. This development was significant not only because it rehabilitated the plausibility of memory transfer, but also because it indicated a mechanism by which transfer could occur under natural conditions, namely by packaging RNA into virus-like particles (``capsids'').

This mechanism may extend far beyond transfer of bacterial pathogen avoidance. Around 8\% of the human genome consists of endogenous retroviruses---genetic material that was originally inserted into germline DNA by a virus and then passed to offspring \citep{lander01}. These genes have been repurposed to serve a variety of functions, one of which is the construction of capsids for packaging of RNA and possibly other molecules. Capsids can be transported out of the cell by by extracellular vesicles, endocytosed in other cells, and disassembled. Messenger RNA contained in the capsids can then be translated into protein. In principle, non-coding RNA transmitted through capsids could play a regulatory role in other cells. This pathway constitutes a phylogenetically ancient form of intercellular communication \citep{pastuzyn18,hantak21}. If memories are stored in RNA, this communication pathway provides a mechanism by which other neurons could share durably encoded messages.

It is tempting to speculate that intercellular transmission of RNA is a relic of the ``RNA world'' that likely preceded the ``DNA world'' in which we currently live \citep{gilbert86}. In the RNA world, genetic information was stored in RNA rather than DNA. Although this storage format is inherently less stable, it might still have been sufficient to support simple lifeforms (and continues to serve as the genetic material for some viruses). After the emergence of DNA, the information storage capabilities of RNA may have been utilized by cells to encode non-genetic information---possibly one function (among many) implemented by non-coding RNAs \citep{mattick06,akhlaghpour20}.

\subsection{Nuclear mechanisms: DNA methylation and histone modification}

Molecular memory theorists in the 1960s did not yet know about the complex mechanisms for gene regulation that exist in the nucleus. Most importantly for our purposes, some of these mechanisms are modifiable by experience and can even be inherited transgenerationally (hence the designation ``epigenetic''). While my focus is not on transgenerational inheritance, I think it is unlikely to be a coincidence that these mechanisms also play a role in memory storage within a single organism's lifetime. As mentioned earlier, it seems plausible that if evolution discovered a mechanism for transgenerational memory storage, this mechanism could be repurposed for memory storage.

In a prescient discussion of stable molecular storage, \citet{crick84} was inspired by the mechanism of DNA methylation to illustrate how a bistable swich could store information despite molecular turnover. The basic unit is a protein dimer whose monomers are modifiable in a binary manner by an enzyme (e.g., phosphorylation by a kinase, or methylation by methyltransferase). In principle, the dimer can be in one of 4 possible states, but Crick further suggested that enzymatic modification of a monomer would only occur if the other monomer was already modified. Thus, the dimer would persistently be in one of two states, ``active'' (+,+) and ``inactive'' (-,-). If one of the monomers were to lose its modification state (e.g., through protein degradation), the enzymatic process would repair it.

Crick's arguments were very general; he did not seriously consider DNA methylation itself as potential storage site. \citet{holliday99} addressed this possibility more directly. Because methylation of gene promoters typically suppresses transcription, the methylation state of DNA regulates gene expression and thereby controls cellular signaling. Holliday pointed out that the potential storage capacity of DNA methylation is stupendously large: if we think of the methylation state as a binary code, then even just 30 methylation sites could represent more than a billion possible codewords, and thus the millions of sites in the human genome could store a vast number of memories.

It is now well-known that DNA methylation is affected by the history of a cell's inputs, that these changes can sometimes be long-lasting, and that they affect animal behavior \citep{day10}. For example, \citet{miller10} found that contextual fear conditioning resulted in methylation of a memory-associated gene (calcineurin) in the dorsomedial prefrontal cortex, which persisted 30 days later. Furthermore, memory expression was reduced by inhibition of DNA methyltransferase 30 days after training.

Another epigenetic mechanism that has received considerable attention in the cell biology literature is histone modification. DNA is packaged into a compact chromatin structure by binding to histone proteins, which protect the DNA from damage and prevent tangling. In addition, histones regulate gene expression, for example by controlling the access of transcription factors to DNA. It has been suggested that the pattern of posttranslational histone modifications functions as a code that stores information about a cell's history \citep{jenuwein01,turner02}.

The study of histones in the brain has a long history, going back to early evidence that long-term memory is associated with histone acetylation \citep{schmitt79}, a finding corroborated by later studies \citep{swank01,levenson04}. Disruption of histone acetylation, either by genetic knockout or by pharmacological inhibition of histone deacetylase, impairs long-term memory \citep{alarcon04,korzus04,levenson04,stefanko09}. Remarkably, inhibition of histone deacetylase can even restore access to memories after severe brain atrophy \citep{fischer07}.

In summary, there exist nuclear mechanisms that can plausibly function as memory storage substrates, and which have been repeatedly implicated in behaviorally expressed memory. The major gap in our understanding is the nature of the memory code---the \emph{codebook}, in the language of communication systems. Despite liberal use of the word ``code'' in the cell biology literature, it is still unclear what exactly is being encoded and how. The situation in neurobiology is even more challenging, because most researchers in that area do not even regard these nuclear mechanisms as codes at all. So deeply entrenched is the synaptic view of memory that these mechanisms are widely regarded as playing a supportive role for synaptic plasticity and maintenance \citep{zovkic13}. The idea that they could autonomously function as long-term storage substrates has only recently been pursued \citep[see][]{gold21}.

\section{A new theoretical perspective}

Given the empirical and conceptual problems with the classical synaptic plasticity theory, we are left with the challenge of repairing the theory and bridging the gap with non-classical molecular mechanisms. I will approach this challenge in three parts, following the template of Marr's levels \citep{marr82}: (1) The computational problem; (2) the algorithmic solution; and (3) the physical implementation. At present, this theory is just a sketch---a starting point for a new direction of thinking.

\subsection{Two computational problems: inference and learning}

Peripheral receptors in the nervous system collect sensory observations, and it is the job of the brain to ``make sense'' of these observations. Bayesian theories of the brain hold that this sense-making process corresponds to probabilistic inference over the latent causes of observations \citep{knill04}. As already mentioned, Bayesian inference is not tractable in general, and different theories make different claims about how inference is approximated \citep{gershman17}. The intractability of Bayesian inference also creates a problem for learning the generative model, because computing the computing the posterior over parameters also involves the same intractable marginalization.

To formalize these points, let $\mathbf{x}$ denote the observations, $\mathbf{z}$ the latent causes, and $\theta$ a set of parameters that govern the generative model $p(\mathbf{x},\mathbf{z},\theta)$, a joint probability distribution over all of the variables. To appreciate the generative flavor of this model, it is useful to decompose the joint distribution into the product of factors that define a causal process:
\begin{align}
    p(\mathbf{x},\mathbf{z},\theta) = p(\mathbf{x}|\mathbf{z},\theta) p(\mathbf{z}|\theta) p(\theta).
\end{align}
Reading from right to left, this equation says that we can draw samples from the generative model in a sequence of steps: (i) draw parameters $\theta$; (ii) draw latents $\mathbf{z}$ conditional on the parameters; (iii) draw obeservations $\mathbf{x}$ conditional on the latents and parameters.

After observing $\mathbf{x}$, Bayes' rule stipulates how to ``invert'' the generative model, producing a posterior distribution over $\mathbf{z}$ and $\theta$:
\begin{align}
    p(\mathbf{z},\theta|\mathbf{x}) = \frac{p(\mathbf{x},\mathbf{z},\theta)}{p(\mathbf{x})}.
    \label{eq:bayes}
\end{align}
The denominator (the marginal likelihood) requires marginalizing the joint distribution over $\mathbf{z}$ and $\theta$, which is computationally intractable in general. Thus, we cannot expect the brain to carry out exact Bayesian inference for arbitrary generative models.

To facilitate the algorithmic solution I present in the next section, it will be useful to reformulate the computational problem in a different way. For any distribution of the form $q(\mathbf{z},\theta|\mathbf{x})$, the following equality holds:
\begin{align}
    \log p(\mathbf{x}) = \mathcal{D} - \mathcal{F},
\end{align}
where
\begin{align}
    \mathcal{D} = \mathbb{E}_q\left[\log \frac{q(\mathbf{z},\theta|\mathbf{x})}{p(\mathbf{z},\theta|\mathbf{x})} \right]
\end{align}
is the \emph{Kullback-Leibler (KL) divergence} between $q(\mathbf{z},\theta|\mathbf{x})$ and the posterior $p(\mathbf{z},\theta|\mathbf{x})$, and
\begin{align}
    \mathcal{F} = \mathbb{E}_q\left[ \log \frac{q(\mathbf{z},\theta|\mathbf{x})}{p(\mathbf{x},\mathbf{z},\theta)} \right]
\end{align}
is the \emph{Helmholtz free energy}. The notation $\mathbb{E}_q[\cdot]$ denotes an expectation with respect to $q(\mathbf{z},\theta|\mathbf{x})$. Because the KL divergence is always non-negative, $-\mathcal{F}$ is a lower bound on the log marginal likelihood $\log p(\mathbf{x})$.\footnote{In statistical machine learning, $\log p(\mathbf{x})$ is sometimes referred to as the \emph{evidence} and hence $-\mathcal{F}$ as the \emph{evidence lower bound}.} The KL divergence is minimized to 0 when $q(\mathbf{z},\theta|\mathbf{x}) = p(\mathbf{z},\theta|\mathbf{x})$. Thus, exact Bayesian inference is equivalent to finding a conditional distribution that minimizes free energy. This is the key idea underlying variational inference algorithms \citep{wainwright08}, and serves as the foundation of the free energy principle in neuroscience \citep{friston10}.

We haven't yet bought anything algorithmically by formulating Bayesian inference as an optimization problem, since minimizing KL divergence to 0 requires that we have access to the true posterior. There are two reasons to adopt this formulation. One is conceptual: it allows us to think about inference and learning as optimizing the same objective function \citep{neal98}. The other reason, which I pursue in the next section, is that it allows us to derive tractable algorithms by restricting the conditional distribution in some way.

\subsection{An algorithmic solution}

One way to make free energy minimization tractable is to restrict the variational posterior to a class of distributions (inference models) that are differentiable with respect to their parameters, which enables the use of gradient-based techniques for optimization. This is the key idea in \emph{neural variational inference} \citep{mnih14}, which uses a neural network, parametrized by $\phi$, as the inference model $q_\phi(\mathbf{z}|\mathbf{x})$.\footnote{Variational autoencoders \citep{kingma13} follow essentially the same idea, but use a different estimator of the gradient.} This technique effectively ``amortizes'' the posterior computation by replacing Bayes' rule (which takes as input both observations and a generative model) and replaces it with a function that takes as input only the observations \citep{stuhlmuller13,gershman14}. This can dramatically reduce the computational cost of inference, at the expense of flexibility.

\subsubsection{Minimizing free energy by gradient descent}

The gradient of the free energy with respect to $\phi$ is given by:
\begin{align}
    \nabla_\phi \mathcal{F} = \mathbb{E}_q\left[ \nabla_\phi q_\phi(\mathbf{z}|\mathbf{x}) \log \frac{q_\phi(\mathbf{z}|\mathbf{x})}{p(\mathbf{x},\mathbf{z},\theta)} \right].
    \label{eq:gradphi}
\end{align}
We can apply the same idea to learning the parameters $\theta$, using an inference model $q_\lambda(\theta|\mathbf{x})$ parametrized by $\lambda$ and taking the gradient of the free energy with respect to $\lambda$:
\begin{align}
    \nabla_\lambda \mathcal{F} = \mathbb{E}_q\left[ \nabla_\lambda q_\lambda(\theta|\mathbf{x}) \log \frac{q_\lambda(\theta|\mathbf{x})}{p(\mathbf{x},\mathbf{z},\theta)} \right].
\end{align}
The separation of updates for $\phi$ and $\lambda$ implies a factorized (or ``mean-field'') posterior approximation:
\begin{align}
    q(\mathbf{z},\theta|\mathbf{x}) = q_\phi(\mathbf{z}|\mathbf{x}) q_\lambda(\theta|\mathbf{x}).
\end{align}
The marginalization over $\mathbf{z}$ can be approximated by drawing $N$ samples from $q(\mathbf{z},\theta|\mathbf{x})$:
\begin{align}
    &\nabla_\phi \mathcal{F} \approx \frac{1}{N} \sum_{n} \nabla_\phi q_\phi(\mathbf{z}^n|\mathbf{x}) \log \frac{q_\phi(\mathbf{z}^n|\mathbf{x})}{p(\mathbf{x},\mathbf{z}^n,\theta^n)},
\end{align}
where $\mathbf{z}^n$ and $\theta^n$ denote posterior samples. An analogous expression applies to $\nabla_\lambda \mathcal{F}$.

Using these gradients, approximate inference and learning can be carried out by stochastic gradient descent (the ``stochastic'' part refers to the fact that the gradient is being approximated using samples). The simplest form of stochastic gradient descent is given by:
\begin{align}
    &\Delta \phi = -\alpha_\phi \nabla_\phi \mathcal{F} 
    \label{eq:update1} \\
    &\Delta \lambda = -\alpha_\lambda \nabla_\lambda \mathcal{F}
    \label{eq:update2}
\end{align}
where $\alpha_\phi$ and $\alpha_\lambda$ are learning rates.

This algorithmic solution provides high-level answers to the content and structure questions about memory. Content consists of the variational parameters $\phi$ and $\lambda$. From the perspective of free energy optimization, there is no substantive difference between these parameters. However, from the perspective of biology, there is a substantive difference: these parameters may be encoded by different substrates, one synaptic ($\phi$) and one molecular ($\lambda$). The structure of memory consists of writing and reading operations for these two types of content. Memory is written via the variational update equations (Eqs. \ref{eq:update1} and \ref{eq:update2}). Memory is read via sampling $q_\phi$ and $q_\lambda$. Because of their qualitatively different substrates, I hypothesize qualitatively different neural mechanisms for these operations.

\subsubsection{Informativeness and associability}

A complete algorithmic theory of learning and memory should explain why learning proceeds more quickly under some conditions than others. In animal learning theory, this has traditionally been discussed in terms of ``associability''---the speed at which two stimuli enter into association. Associability is commonly formalized as a learning rate parameter in models of associative learning \citep[e.g.,][]{bush55,rescorla72,mackintosh75}. In the context of Pavlovian conditioning, we can define associability more broadly as the amount of training required to produce some criterion level of conditioned responding. Balsam, Gallistel, and their colleagues have argued that one cannot understand this broader notion of associability in associative terms \citep{balsam06,balsam09,ward12,ward13}. This argument goes hand-in-hand with their critique of synaptic plasticity as the vehicle for learning: if synaptic plasticity is associative and learning is not, then synaptic plasticity cannot implement learning.

As an alternative, Balsam and colleagues have formalized associability as the mutual information between the conditioned stimulus (CS) and the timing of the unconditioned stimulus (US). They refer to this quantity as the \emph{informativeness} of the CS. Here I will briefly relate this conceptualization to the theoretical framework described above.

In the setting of Pavlovian conditioning, the latent cause is a scalar ($z$) representing the time at which the US will occur. On average, the gradient of the free energy is given by:
\begin{align}
    \mathbb{E}_p[\nabla_\phi \mathcal{F}] = \mathbb{E}_{q,p}\left[ \nabla_\phi q_\phi(z|\mathbf{x}) \log \frac{q_\phi(z|\mathbf{x})}{p(z|\theta)} \right] + \text{const.}
\end{align}
where $\mathbb{E}_p[\cdot]$ denotes an expectation with respect to $p(\mathbf{x})$, and $\mathbb{E}_{q,p}[\cdot]$ denotes an expectation with respect to $q_\phi(z|\mathbf{x})p(\mathbf{x})$. If we further assume that the gradient of the variational posterior is approximately constant across $\mathbf{x}$ and $z$, $\nabla_\phi q_\phi(z|\mathbf{x}) \approx c(\phi)$, we obtain:
\begin{align}
    \mathbb{E}_p[\nabla_\phi \mathcal{F}] \approx  c(\phi) \cdot \mathbb{E}_{q,p}\left[ \log \frac{q_\phi(z|\mathbf{x})}{p(z|\theta)} \right] + \text{const.}
\end{align}
The right-hand-side (ignoring the constants) is proportional to the mutual information between the CS configuration ($\mathbf{x}$) and US time ($z$). I have thus established conditions under which learning rate in the free energy framework approximately tracks informativeness, a probabilistic formalization of associability.

\subsection{Biologically plausible neural implementation}

The algorithmic solution of the preceding section suggests that we should look for two types of memories in the brain. I argue here that these types are subserved by qualitatively different cellular systems: inference parameters are stored at the synapse and updated via synaptic plasticity, while generative parameters are stored in a molecular format within the cytoplasm or nucleus and updated via biochemical processes.

\subsubsection{Circuit model}

I hypothesize a ``direct coding'' scheme in which the spiking of individual neurons reports a random sample of a variable conditional on the neuron's input. This kind of scheme has been widely used in neural models of probabilistic inference \citep[e.g.,][]{pecevski11,haefner16,hoyer03,buesing11,orban16}.\footnote{Two notable alternatives to direct coding are predictive coding, in which a neuron reports a prediction error \citep{bastos12,friston05,deneve08}, and probabilistic population coding \citep{ma06}, in which a variable is represented by the spiking activity of a neural population.} For ease of exposition, I will assume that all variables are binary. Let $\rho_i$ denote the firing rate of cell $i$, which I take to report the posterior probability that $z_i=1$. These ``latent cause'' cells receive input from a separate population of ``observation'' cells reporting the occurrence of sensory variables as well as from a subset of other latent cause cells (see below). I use $\mathbf{v}_i$ to denote the vector of all inputs to cell $i$.

The implementation of the posterior over latent causes is a neural network in which firing is driven by a Poisson process with rate $\rho_i$ \citep[see also][]{rezende14}. I assume that the firing rate is an exponential function of the membrane potential $u_i$,
\begin{align}
\rho_i = \exp(u_i - \psi_i),    
\end{align}
with threshold $\psi_i$ \citep{jolivet06}. The membrane potential is a linear function of synaptic inputs:
\begin{align}
    u_i = \sum_{j} w_{ij} v_{ij},
\end{align}
where $j$ indexes synaptic inputs to cell $i$, and $w_{ij}$ is the synaptic strength. The firing rate $\rho_i$ corresponds to the inference model component $q_\phi(z_i=1|\mathbf{v}_i)$, and the synaptic strengths $\mathbf{W}$ and the thresholds $\psi$ correspond to the inference parameters $\phi$.

I propose that the posterior over generative parameters, $q_\lambda(\theta|\mathbf{x})$, is not expressed through spiking but rather through gene expression. Samples of $\theta$ correspond to RNA transcripts which specify the instructions for evaluating the joint probability of the random variables (see below). I refer to this process as ``querying'' the generative model. The RNA samples are controlled by a transcription program, parametrized by $\lambda$. For example, $\lambda$ might correspond to nuclear marks such as histone acetylation or DNA methylation states.

\subsubsection{Connectivity}

To map the spiking neuron circuit onto the inference model, we need to specify the dependency structure of the inference model. I assume that $q_\phi$ factorizes into a set of conditional distributions, one for each latent cause. To ensure that this factorization defines a valid distribution, we require that the connectivity structure of the latent cause cells corresponds to a directed acyclic graph (i.e., no loops between cells). We can then represent the approximate posterior over $\mathbf{z}$ as follows:
\begin{align}
    q_\phi(\mathbf{z}|\mathbf{x}) = \prod_i q_\phi(z_i|\mathbf{x},\mathbf{z}_{\text{pa}(i)}),
\end{align}
where $\text{pa}(i)$ denotes the parents of latent cause $z_i$.

To ensure locality of the plasticity rules given below, each cell needs to receive additional inputs. Let $\mathcal{C}$ denote a partition of the variable set $\mathbf{h} = \{\mathbf{x},\mathbf{z},\theta\}$ such that each subset contains at most one latent cause ($z_i$) and the other variables in the subset belong to the Markov blanket of $z_i$. The joint distribution can then be expressed as the product of component distributions:
\begin{align}
    p(\mathbf{x},\mathbf{z},\theta) = p(\mathbf{h}) = \prod_{c \in \mathcal{C}} p_c(\mathbf{h}_c|\mathbf{h}_{\text{pa}(c)}),
\end{align}
where pa$(c)$ denotes the parents of $\mathbf{h}_c$. Note that this factorization can be completely different from the inference model factorization.

Each component of the partition maps to a single latent cause (i.e., $i$ and $c$ are in one-to-one correspondence). This means that each cell $i$ should receive inputs from all the variables in the union of $c$ and $\text{pa}(c)$. This poses a challenge, because these inputs cannot depend on the synaptic parameters; the cell needs to be able to sense the values of these inputs independently of the inference parameters. One way to finesse this issue is to utilize the concept of a postsynaptically \emph{silent synapse}, which lacks AMPA receptors (and hence does not respond to moderate levels of depolarization) but possesses NMDA receptors \citep{malenka97}.\footnote{Alternatively, the synapse may possess both AMPA and NMDA receptors, but a sufficiently small glutamate concentration would bind preferentially to NMDA receptors, due to their substantially higher affinity.} If, as conventionally assumed, synaptic plasticity is mediated by trafficking of AMPA receptors to the postsynaptic membrane, then an inference model that depends on AMPA receptors can be separated from a generative model that depends on NMDA receptors. Specifically, a cell could receive inputs about its probabilistic dependencies through NMDA receptors without eliciting AMPA currents that primarily control spiking activity (i.e., sampling from the inference model).

\subsubsection{Plasticity rules}

The weight parameters can be updated using a form of three-factor synaptic plasticity that is equivalent to a sample-based approximation of the gradient (Eq. \ref{eq:gradphi}):
\begin{align}
    \Delta w_{ij} = \alpha_\phi \underbrace{v_j}_{\text{pre}} \underbrace{(\hat{z}_i - \rho_i)}_{\text{post}}(L_i - \hat{z}_i\log \rho_i + \rho_i),
\end{align}
where $\hat{z}_i$ denotes the activity state (0 or 1) of the latent cause cell. Similarly, the threshold can be updated according to:
\begin{align}
    \Delta \psi_i = - \alpha_\phi (\hat{z}_i - \rho_i)(L_i - \hat{z}_i\log \rho_i + \rho_i).
\end{align}
The term $L_i = \log p(\mathbf{v},\hat{z}_i,\theta)$ is a signal that I argue is generated by querying the intracellular generative model (discussed further below).

The plasticity rule for the inference parameters takes an analogous form:
\begin{align}
    \Delta \lambda = \alpha_\lambda \nabla_\lambda q_\lambda(\theta|\mathbf{x}) (L_i - \hat{z}_i\log \rho_i + \rho_i).
\end{align}
The key difference is that this rule depends on the gradient of the variational posterior over generative parameters with respect to the inference parameters $\lambda$. Recall that $\lambda$ is implemented as a set of transcriptional parameters (e.g., histone or DNA methylation marks) in the nucleus. Thus, the model requires that the transcription probability mass function is differentiable with respect to these marks. Since the marks are typically taken to be binary, it's not immediately clear how this would work, but it's possible that the transcription parameters correspond to mark \emph{frequencies} (i.e., the proportion of marks in a particular binary state), which would then make the transcription probability mass function differentiable.

\subsubsection{Querying the generative model}

Querying the generative model consists of evaluating the probability of $\mathbf{h}$ under the joint probability distribution parametrized by $\theta$. If RNA encodes $\theta$, then one possibility is that signals encoding $\mathbf{x}$ and $\mathbf{z}$ bind to RNA, and then a specialized molecule reads the bound complex to report the scalar log joint probability $L_i$, making it accessible to the plasticity rules described above.

\subsection{Summary and implications}

The theoretical framework presented here attempts to unify synaptic and molecular memory mechanisms under the authority of a single optimization principle. Synaptic learning optimizes an approximate posterior over latent causes; synapses are the storage sites of the parameters governing this posterior. Intracellular/molecular learning optimizes an approximate posterior over generative parameters; nuclei are the storage sites of the parameters governing this posterior.

This synthesis of mechanisms can potentially explain the disjunction between properties of synaptic plasticity and properties of behavioral learning reviewed above. Learning ``facts'' about the environment does not require synaptic plasticity according to my account (facts are stored inside the cell), but synapses are necessary for the expression of memory for these facts, and synaptic plasticity exists to optimize this expression. This is why synaptic plasticity accompanies learning, and why disruption of synaptic plasticity interferes with the expression of memory, even though synapses may not be the sites of memory storage.

My account also explains why memories can be transferred (both horizontally and vertically), and why they can survive dramatic brain remodeling (e.g., during metamorphosis or hibernation). When synapses are destroyed, the generative parameters are preserved in the RNA codes, which can circulate between cells in virus-like capsids \citep{hantak21}. As long as the cells contain generic programs for reading the information stored in RNA, they are not dependent on a particular synaptic circuit. On the other hand, these memories may become less acessible after synapse elimination or remodeling. The temporary loss of accessibility, which may be restored under certain circumstances, is broadly consistent with studies of experimental amnesia reviewed earlier, which showed that memory loss due to amnestic agents (e.g., protein synthesis inhibitors) can be temporary.

\section{Insights from cognitive science}

While I have focused on biological considerations, the idea of amortized inference is also relevant for understanding a number of puzzling observations in the cognitive science literature. Here I briefly review some of the evidence and its theoretical implications.

I begin by noting that Bayes' rule is purely syntactic, in the sense that it can be applied to \emph{any} joint distribution over observations and latent causes. Interestingly, however, human inference is not purely syntactic: probability judgments are sensitive to semantic properties of the joint distribution. In particular, probability judgments are closer to the Bayesian ideal when the joint distribution is ``believable'' (i.e., similar to frequently encountered distributions). \citet{cohen17} asked people to make inferences about medical conditions after being given information about the results of a diagnostic test. They found that people diverged substantially from Bayes' rule when the probabilities were unbelievable. For example, a diagnostic test with a false positive rate of 80\% would be considered unbelievable (no such test would ever be used in the real world). Likewise, a base rate of 50\% for pneumonia would be considred unbelievable (it's not the case that every other person you meet has had pneumonia). The believability effect documented by Cohen and colleagues is reminiscent of a similar semantic bias in syllogistic reasoning, where beliefs about the plausibility of statements influence truth judgments, violating the purely syntactic principles of logic \citep{revlin80,wilkins29,evans83}.

Belief bias is a natural consequence of amortized inference \citep{dasgupta20}. If the function (e.g., a neural network) used to approximate Bayesian inference has limited capacity, then it inevitably has to concentrate this capacity on distributions it encounters frequently. \citet{dasgupta20} tested this hypothesis directly by giving people inference problems drawn from particular distributions. When presented repeatedly with highly diagnostic evidence and low diagnostic base rates, people learned to effectively ignore the base rates, leading to errors when subsequently given test problems where the base rates are highly diagnostic. When presented repeatedly with low diagnostic evidence and high diagnostic base rates, the opposite pattern occurred, with errors on test problems where the evidence was highly diagnostic. Importantly, people behaved differently on the \emph{same} test trials depending on what distribution of inference problems they were given. Other studies have shown that people exhibit serial dependencies between their inferences on problems presented in sequence, even though the problems were strictly independent \citep{dasgupta18}. These serial dependencies may be related to widely observed ``carryover'' effects in public opinion surveys, where answers depend on the order in which questions are asked \citep{tourangeau89,moore02}. Taken together, these findings suggest that people do not apply a fixed, purely syntactic inference algorithm; rather, they \emph{learn to infer}.\footnote{See \citet{wang13} and \citet{wang14} for an alternative explanation of some question order effects in terms of quantum probability theory.} I conjecture that this learning process is implemented by synaptic plasticity.

Learning to infer is part of a broader pattern across cognitive science. In many domains, memory is used not only to store information about the world but also to store information about how to think \citep{dasgupta21}. A person may know the rules of chess or the axioms of mathematics, but may still not be a particularly good chess player or mathematician. These are skills that are acquired from experience, but the nature of this experience is not about ``fact learning'' in the traditional sense of observing the world, since the relevant knowledge is acquired by \emph{thinking more}. We are, so to speak, \emph{learning to think}.

A few examples will illustrate how ubiquitous this kind of learning is \citep[see][for a more extensive review]{dasgupta21}. When children first learn arithmetic, they rely heavily on counting, such that the amount of time it takes to add two numbers is proportional to the smaller of the two numbers \citep{groen72}. By the age of 10, children can answer some addition problems by retrieving the solution from memory, resulting in a much flatter response time function \citep{ashcraft82}. The transition from counting to memory retrieval is best understood as a kind of cognitive strategy learning rather than fact learning. The facts in this case are constructed by computation, and the results of these computations are stored in memory. A similar flattening of response time functions has been observed in mental rotation \citep{jolicoeur85,tarr89}. Initially, the time it takes to recognize an object depends on the angular distance between the object and a canonical orientation. With repeated practice, people are eventually able to retrieve the outputs of mental rotation directly from memory, without mentally rotating the object through intermediate orientations.

In summary, learning to think is conceptually distinct from, and complementary to, fact learning. Gallistel has stressed the importance of fact learning for understanding the biological basis of memory \citep{gallistel17,gallistel21}. I wholeheartedly agree with his argument. At the same time, learning to think may be just as important. It may provide an answer to the puzzle of synaptic plasticity posed in the beginning of this paper.

\section{Conclusions}

To recapitulate the central arguments:
\begin{enumerate}
    \item The available evidence makes it extremely unlikely that synapses are the site of long-term memory storage for representational content (i.e., memory for ``facts'' about quantities like space, time, and number).
    \item Fact memories, or more generally probabilistic beliefs about facts, are plausibly stored in an intracellular molecular format.
    \item Synapses may be the site of long-term memory storage for computational parameters that facilitate fast belief updating and communication.
    \item These two forms of memory (representational and computational) work synergistically to optimize a common objective function (free energy). At the biological level, this synergy is realized by interactions between synaptic and molecular processes within a cell.
\end{enumerate}
Much of this theory remains speculative. Why can so little can be firmly asserted, despite decades of intensive research? One reason is that the associative-synaptic conceptualization of memory is so deeply embedded in the thinking of neurobiologists that it is routinely viewed as an axiom rather than as a hypothesis. Consequently, the most decisive experiments have yet to be carried out.

Looking at the history of research on nuclear mechanisms of memory is instructive because it has long been clear to neurobiologists that DNA methylation and histone modification play an important role. However, the prevailing view has been that these mechanisms support long-term storage at synapses, mainly by controlling gene transcription necessary for the production of plasticity-related proteins \citep{zovkic13}. The notion that these nuclear mechanisms might themselves be storage sites was effectively invisible because such a notion was incompatible with the associative-synaptic conceptualization. It wasn't the case that a non-associative hypothesis was considered and then rejected; it was never considered at all, presumably because no one could imagine what that would look like. This story testifies to the power of theory, even when implicit, to determine how we interpret experimental data and ultimately what experiments we do \citep{gershman21b}.

We need new beginnings. Are we prepared to take a step off the terra firma of synaptic plasticity and venture into the terra incognita of a molecular memory code?

\subsection*{Acknowledgments}

The ideas presented here have benefited enormously from conversations with Randy Gallistel, Jeremy Gunawardena, David Glanzman, Johannes Bill, Anatole Gershman, Michael Levin, and Andrew Bolton. I am also grateful to Tony Zador, Aaron Milstein, Sadegh Nabavi, and Honi Sanders for comments on an earlier draft. This material is based upon work supported by the Center for Brains, Minds and Machines (CBMM), funded by NSF STC award CCF-1231216.

\bibliographystyle{apalike}

\end{document}